# Excitation of spin wave modes of a magnetic dot existing in a shifted vortex-state by an out-of-plane oscillating magnetic field

Yan Liu[1,2], Yan Zhou[1*], P. Dürrenfeld[3], Y. Yin[4], D. W. Wang[5], A. N. Slavin[6] and A. Ruotolo[7]

1. Department of Physics, The University of Hong Kong, Hong Kong, China
2. College of Sciences, Northeastern University, Shenyang 110819, People's Republic of China
3. School of Electronic Science and Engineering and Collaborative Innovation Center of Advanced Microstructures, Nanjing University, Nanjing, China
4. Department of Physics, Southeast University, 211 189 Nanjing, China
5. School of Physics and Electronics, Central South University, Changsha 410083, Hunan, P. R. China
6. Department of Physics, Oakland University, Rochester, Michigan 48309, USA
7. Department of Physics and Materials Science, City University of Hong Kong, Hong Kong, China

* correspondence author:yanzhou@hku.hk

**Abstract:**

Excitation of spin wave modes of a vortex-state magnetic dot by an out-of-plane oscillating magnetic field is studied numerically in the presence of a static in-plane magnetic field. It is shown, that the application of the in-plane static field shifts the position of the vortex core and leads to the separate excitation of different azimuthal dipolar spin waves my perpendicular oscillating field. It is also shown that the excited dipolar azimuthal spin waves excite the gyrotropic mode of the vortex core rotation and, depending on the excitation frequency, cause a significant modification (increase or decrease) of the apparent dissipation rate of the gyrotropic mode. The last effect can be explained by the nonlinear parametric interaction between the gyrotropic mode and the dipolar spin wave modes with different azimuthal indices.

The magnetic vortex is a stable magnetic configuration in nano-patterned elements [1-2], which is commonly identified by two parameters: the polarity of the vortex core, which can point up ($p=1$) or down ($p=-1$), and the chirality of the in-plane magnetization, which can be clockwise ($c=-1$) or counter-clockwise ($c=1$). Recently, the precise knowledge of the vortex dynamics becomes more important due to their prospective applications in many spintronics devices, from extremely sensitive magnetic field sensors to spin polarized current-tunable microwave vortex nano-oscillators and vortex magnetic random access memory [2-4].

The vortex-state magnetic disk has two fundamental types of dynamic spin excitations: the low frequency gyrotropic mode involving a spiral-like motion of the vortex core around its equilibrium position and the high-frequency dipolar spin-wave modes involving precession of magnetization in the in-plane part of the magnetic vortex. The typical frequency of the gyrotropic mode is in the range of hundreds of megahertz, depending on the aspect ratio of the magnetic element [5-6]. The dipolar spin wave modes of the plane part of the vortex have eigenfrequencies in the range of several gigahertz and are quantized due to the strong pinning boundary conditions on the dot lateral edges [7-10]. The dipolar spin wave modes are characterized by integer pairs of indices ($n$, $m$), which indicate the number of nodes of the dynamical magnetization component along the radial ($n$) and azimuthal ($m$) directions. Regarding these two types of dynamics, numerous studies have been carried out, including the vortex gyrotropic motion excited by magnetic fields or by spin-polarized currents [11-18], string-like gyrotropic waves [19], azimuthal modes generated by a magnetic field parallel to the dot plane [20-21], and radially symmetric modes generated by magnetic field pulses perpendicular to the dot plane [22-23].

Furthermore, the coupling between the gyrotropic mode of the vortex core rotation and the spin-wave dipolar modes is responsible for many interesting phenomena. For example, *linear* coupling between the gyrotropic mode and the dipolar spin waves lifts the frequency degeneracy of the clockwise (m=1) and counterclockwise (m=-1) azimuthal dipolar spin wave modes, leading to the zero-field frequency splitting of these modes [24-27]. In turn, the selective vortex core reversal can be achieved by taking advantage of the frequency splitting of the azimuthal dipolar modes caused by their interaction with the static dipolar field of the vortex core and even their interference [28-30]. In addition, the resonance of radial spin waves can directly induce ultrafast vortex core reversal [31-33]. However, to the best of our knowledge, no investigations have been performed regarding the gyrotropic motion directly excited by the radial spin waves.

Here we study the influence of high-amplitude dipolar spin waves excited by oscillating perpendicular magnetic fields on the behavior of the gyrotropic mode in vortex-state circular magnetic dots. If the vortex is at rest at the disk center, the perpendicular magnetic fields can only excite the symmetric radial spin modes (m=0). These symmetric spin waves force the vortex core to stay at the disk center, thus only yielding an expansion and compression of the vortex core, and no gyrotropic motion can appear.

Therefore, in our current work we are studying a shifted non-uniform magnetic vortex state, where a static in-plane magnetic field is used to break the symmetry of the state, and an out-of-plane oscillating magnetic field is applied to excite the spin waves.

We performed micromagnetic simulations in a circular-shaped Permalloy nanodot [Fig. 1(a)] of 200 nm in diameter and 5 nm in thickness, by OOMMF code, which employs the Laudau-Liftshitz-Gilbert equation [34]. The ground state for the disk is assumed to be a magnetic vortex with $(p, c) = (1, 1)$. In the simulations, the nanodisk is discretized into cells with a size of $2.5 \times 2.5 \times 5$ nm$^3$. Typical material parameters for permalloy are used: the saturation magnetization $M_s = 8.6 \times 10^5$ A/m, the exchange constant $A = 1.3 \times 10^{-11}$ J/m, the Gilbert damping parameter $\alpha = 0.01$, and no crystalline anisotropy, i.e., $K_1 = 0$.

Many studies have dealt with spin waves excited in a shifted non-uniform magnetic vortex state [35-38]. However, since the excitation fields have been applied in the disk plane, there is little understanding concerning the spin waves excited by perpendicular magnetic fields. Therefore, as our first step we examine the spin-wave modes in a shifted vortex excited by an out-of-plane magnetic field. We first apply an external in-plane field $H_y$ along the $y$ direction. The field is smaller than the magnetic saturation field to make sure that the stable state is still a vortex state. Under the effect of $H_y$ the vortex core is shifted away from the disk center and stabilized at an off-center position on the -$x$ axis after a period of time. Figure 1(b) shows the typical dependence of the vortex core displacement from the disk center as a function of $H_y$.

Then, we apply a square wave pulse, 0.5 ns in duration and 30 mT in strength, perpendicular to the disk plane to excite high frequency spin-wave modes. Fast Fourier transform (FFT) of the average of the z component of the magnetization ($<m_z>$) yield the spin-wave spectra in a wide frequency range, which are shown as a function of the static bias magnetic field $H_y$ in Fig. 1(c).

When the vortex core is at the center of the disk ($H_y=0$), the spectrum shows three spin-wave modes at 8.6, 12.6 and 16.6 GHz. The corresponding spatially dependent FFT amplitudes are shown in the first row of Fig. 1(d). These patterns indicate that the three spin-wave modes correspond to the radial modes with $n=1$, 2, and 3, respectively. Displacement the vortex core induced by the static in-plane field $H_y$ breaks the symmetry and changes the spectrum completely, with some modes showing very strong frequency shifts as a function of $H_y$. We observe different behaviors for the centered vortex state (CV, $H_y<4$ mT), the biased vortex state (BV, $4<H_y<15$ mT), and the edge-vortex state (EV, $H_y>15$ mT). In the CV state, three eigenmodes are observed with their eigenfrequencies being almost independent of $H_y$. The corresponding FFT power spatial distributions of $m_z$ for the three excited modes at $H_y = 1$ mT are shown in Fig. 1(d). The small shift of the vortex core slightly breaks the cylindrical symmetry, but maintains the characteristics of these modes like that at $H_y= 0$. All the three modes can be described as radial spin waves with the indices $n = 1$, 2 and 3, respectively. A further increased in-plane bias field leads to a larger shift of the vortex core. In the BV state, the eigenmodes change more significantly. The $n = 1$ mode splits up into four distinct modes, labeled as p1, p2, p3, and p4. The $n=2$ mode (p5) does not split, but its frequency decreases significantly when $H_y$ increases. The $n=3$ mode splits up into two modes, labeled as p6 and p7. The frequency of the main branch (p7) decreases quickly, while the mode p6 appears at around $H_y = 8$ mT and gradually merges into p7. As examples we show the FFT power spatial distributions of $m_z$ at the frequencies of these excited modes when $H_y = 9$ mT in Fig. 1(d). It is evident that the nodes' symmetry along the radial direction is destroyed, yet still with some characteristics similar to $H_y = 0$. When the bias field is even further increased, the vortex core is close to the disk edge (EV state). In this state, besides the modes p1, p2, p3, p4, and p7, two additional modes appear, labeled as EM-p9 and EM-p10.

As the FFT power spatial distributions of $m_z$ at $H_y$=20 mT in Fig. 1(d) shows, these modes are completely different. The images show the formation of stripes in the region far from the vortex core, which is very similar to the case of a uniform state [39]. Thus, it is not possible to associate these modes with corresponding modes that exist at a zero in-plane bias field. Indeed, previous studies have been mainly focused on the spin waves in the absence of a bias magnetic field when the vortex core was localized at the dot center. The spin-wave modes ($n$, $m$) classification is usually applicable to the zero-field case [7-9]. For azimuthal modes, it was extended up to the vortex nucleation field [37]. However, for our case, we think that such a classification scheme is only applicable to the CV state and the very beginning of the BV state.

At our second step, we investigate the interactions between the vortex gyrotropic motion and the spin-wave modes discussed above. In a first study, we directly induce the gyration of a shifted vortex by the spin waves. In this case, we use the shifted vortex with different $H_y$ as the initial state, and apply a high frequency oscillating magnetic field $H_z = H_0 \sin(2\pi f_z t)$ perpendicular to the disk, where $H_0$ is the field amplitude, and $f_z$ is the field frequency, ranging from 1 to 20 GHz.

We present the $y$ component of the magnetization $<m_y>$ average over the disk as a function of simulation time at different $f_z$ when $H_0$=30 mT and $H_y$=10 mT in Fig. 2(a). Except the GHz-perturbations directly caused by the oscillating $H_z$, the vortex gyrotropic motion can be inferred from the MHz-oscillations of $<m_y>$, which are shown as the lower frequency part of the corresponding Fourier spectra in Fig. 2(b). One also notices that $f_z$ has a strong influence on the vortex gyrotropic motion. The amplitude of $<m_y>$ is very small at 5.0 and 19.0 GHz. That is, the vortex core has no obvious motion. However, the amplitude increases with time when $f_z$ assumes 9.0, 11.2, and 15.6 GHz, indicating the excitation of the vortex gyration. Figure 2(c) shows the trajectory of the vortex core, a spiral motion, for $f_z = 15.6$ GHz. Its distance from the original position increases with time until it reaches a stable value, see Fig. 2(d). As a measure of the amplitude of the gyrotropic motion, we here define the distance of the vortex core from its initial position at $t$ = 20 ns as $d_v$. We also notice that the oscillations of $<m_y>$ are abruptly stopped for $f_z = 7.0$, indicating that the magnetic state of the disk is no longer a vortex. This can be avoided by reducing $H_0$ to 10 mT. Thus, in the following calculations, the value of $H_0$ are chosen to be 10 mT when $f_z < 10$ GHz in order to maintain the vortex state.

To clarify in more detail the relation between the gyrotropic motion of the shifted vortex and the perpendicular oscillating fields, we present the simulated $d_v$ in the two-dimensional parameter space of ($H_y$, $f_z$) in Fig. 2(e). A series of excited peaks appears, which we can also label as p1, p2, .... Interestingly, almost all the excited peaks correspond to the spin-wave modes labeled as the same symbols in Fig. 1(c). Comparing the top panels of Fig. 1(c) and Fig. 2(e), one can see that the marked peak frequencies in the $d_v$ curve agree well with the eigenfrequencies in the spin-wave spectra, except for mode p1. This general agreement confirms that the gyrotropic motion of the shifted vortex can be excited by spin waves, which are induced by the perpendicular oscillating magnetic field.

Finally, we focus on another condition, where the influence of the spin waves on the vortex gyration is studied for the case when the vortex core is gyrating in the disk. Here, we simultaneously apply the bias field $H_y$ and the driving field $H_z$ to the disk with the vortex core initially in the disk center. Under the effect of $H_y$, the vortex core will gyrate around an off-centered equilibrium position, while at the same time the spin waves are excited by $H_z$. Figure 3(a) shows the oscillations of the magnetization $<m_y>$ at the vortex gyration frequency for $H_y$=4 mT. For the case of zero spin waves ($H_z = 0$), the amplitude of $<m_y>$ clearly decreases with the simulation time, which is generally known to be due to the existence of damping. However, when we include the effect of $H_z$, significant changes on the oscillations of $<m_y>$ can be observed. As an example, it shows anti-damping at 10.6, 12.2, and 16.6 GHz, while the damping is enhanced at 11.4 and 15.8 GHz.

To obtain a more generalized understanding, we can calculate the effective damping constant $\alpha'$ for a series of $H_y$ and $f_z$, where the bias field $H_y$ is varied from 0 to 13 mT. As defined in the top panel of Fig. 3(a), the effective damping constant is calculated through $\alpha' = \alpha \times \left( \omega_I(H_y, f_z) / \omega_I(H_y, 0) \right)$, where $\omega_I(H_y, f_z) = (1/\Delta t) \ln(A(t_2)/A(t_1))$. The calculated $\alpha'$ in the plane of ($H_y$, $f_z$) is shown in Fig. 3(b). There are both anti-damping peaks (p2, p3, p4, p5, p7, p8) and peaks with additional damping (a1, a2, a3, a4). The anti-damping peaks are consistent with the spin-wave modes in the CV state, yet deviating largely from the spin-wave modes at large $H_y$. The peaks with additional damping can not be related to any spin-wave mode in Fig. 1(c). This is due to the spin-wave spectrum being different when the vortex core gyrates in the disk. As described in Ref. 30, the eigenfrequencies strongly depend on the amplitude of the gyrotropic mode.

We believe that the possible explanation of the behavior of damping of the gyrotropic mode under the strong excitation of the dipolar spin wave modes by perpendicular oscillating magnetic field can be related to the parametric interaction of the gyrotropic mode with the azimuthal dipolar modes having opposite azimutahl indices $m$. It is known that linear interaction of these modes leads to the frequency splitting between the azimuthal mode [26], and the azimutal mode (m=-1), having the sense of azimuthal rotation opposite to the sense of gyration of the gyrotropic mode, has a higher frequency ( $f_{m=-1} > f_{m=+1}$). If now azimuthal mode with the frequency $f_{m=-1}$ is excited by the perpendicular oscillating field with a sufficiently large amplitude, this mode can experience a parametric decay into the lower frequency azimuthal mode $f_{m=+1}$ and the gyrotropic mode $f_G$: $f_{m=-1} \to f_{m=+1} + f_G$. This decay process will have an power threshold determined by the magnetic damping in the dot material and will lead to the transfer of energy from the mode $f_{m=-1}$ linearly excited by the perpendicular oscillating field to the gyrotropic mode $f_G$, and, therefore, to the apparent decrease of damping of this mode.

On the other hand, if the lower frequency azimuthal dipolar mode $f_{m=+1}$ is excited by the perpendicular oscillating field, this mode could participate in the parametric confluence process $f_{m=+1} + f_G \to f_{m=-1}$, which does not have a power threshold. This process will lead to the energy transfer from the gyrotropic mode to the higher frequency azimuthal dipolar mode $f_{m=-1}$, and, therefore, to the apparent increase of dissipation of the gyrotropic mode $f_G$. Further numerical simulations are necessary to confirm our hypothesis about the resonance parametric interaction between the dipolar and gyrotropic modes in a magnetic dot existing a shfted vortex state.

In conclusion, we present a numerical evidence about the nonlinear coupling between the vortex gyrotropic motion and higher frequency dipolar spin wave modes in a magnetic dot containing a shifted vortex. The spin waves excited by an out-of-plane oscillating magnetic field not only can induce a shifted vortex to gyrate, but also can change the apparent gyration damping when the vortex is gyrating in the disk, which indicated the nonlinear character of the mode coupling leading to the energy exchange between the modes.

## Acknowledgments

Y.Z. acknowledges the support by the Seed Funding Program for Basic Research and Seed Funding Program for Applied Research from the HKU, ITF Tier 3 funding (ITS/203/14), the RGC-GRF under Grant HKU 17210014, and University Grants Committee of Hong Kong (Contract No. AoE/P-04/08). This work was supported in part by the National Natural Science Foundation of China (Grant No. 11404053).

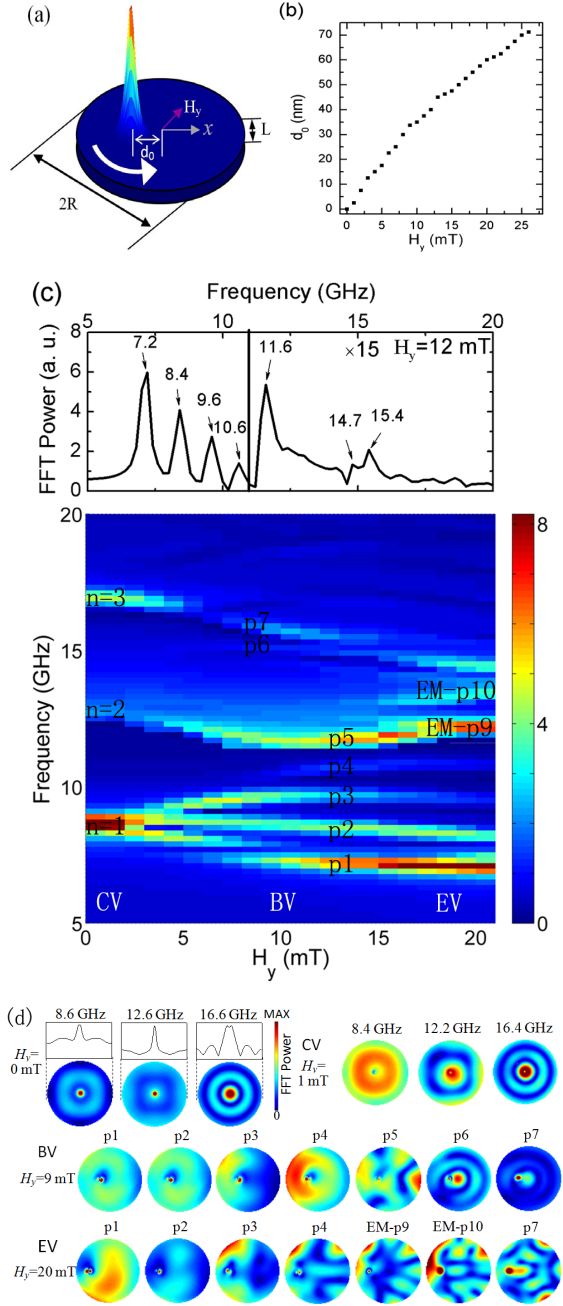

Fig. 1 (a) Schematic illustration of the nanodisk and its shifted vortex state, wherein vortex-core magnetization is upward and in-plane curling magnetization is counter-clockwise (white arrow). (b) Vortex core displacement from the disk center ($d_0$) as a function of $H_y$. (c) FFT power plots of the spin-wave spectra and its evolution as a function of $H_y$. The symbols p1, p2... label the spin-wave eigenmodes discussed in the text. Note that, for a better look, the values of FFT power for $f > 11$ GHz are magnified 15 times. A single spin-wave spectra like that for $H_y=12$ mT is shown on the top, where the excited peaks and their eigenfrequencies are marked. (d) FFT power spatial distributions of $m_z$ for the spin-wave modes at $H_y=0$, 1 (CV), 9 (BV), and 20 mT (EV).

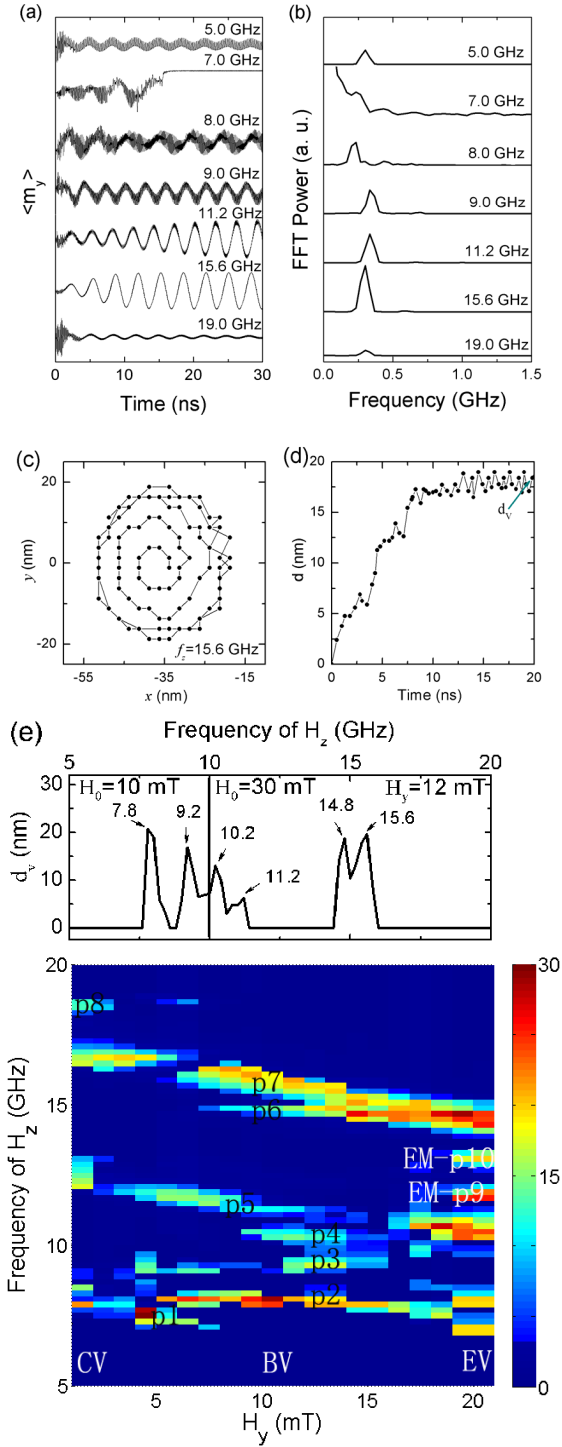

Fig. 2 Spin waves-induced gyration of a shifted vortex. (a) Variation of $<m_y>$ of the shifted vortex ($H_y$=10 mT) as a function of simulation time when excited by a perpendicular oscillating magnetic field with the amplitude $H_0$=30 mT, where $f_z$=5.0, 7.0, 8.0, 9.0, 11.2, 15.6 and 19.0 GHz, respectively. (b) The corresponding FFT power spectrum of $<m_y>$ given in (a). (c) The trajectory of vortex core, and (d) its corresponding distance from its original position (*d*) for $f_z = 15.6$ GHz. (e) Color plot of $d_v$ with respect to $H_y$ and $f_z$. A single $d_v$ curve like that for $H_y$=12 mT is shown on the top, where the excited peaks and their frequencies are marked.

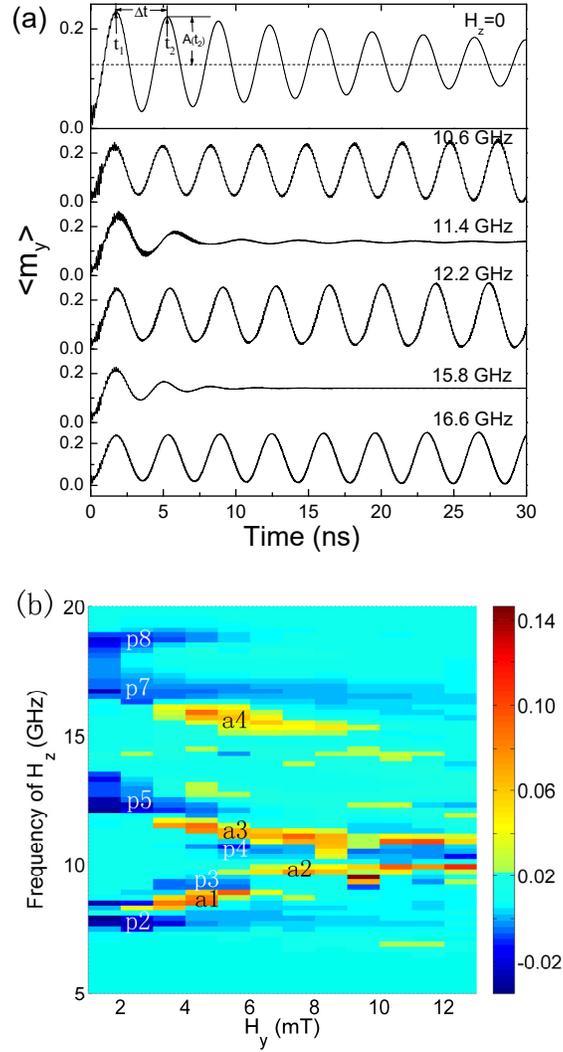

Fig. 3 Vortex gyration mediated by spin waves. (a) Variation of $<m_y>$ as a function of simulation time excited by a bias field $H_y$=4 mT when $H_z$=0, and $H_z$=30 mT with $f_z$=10.6, 11.4, 12.2, 15.8, and 16.6 GHz, respectively. (b) Color plot of the effective damping constant $\alpha$ with respect to $H_y$ and $f_z$. The symbols p1, p2… label the anti-damping modes, and the symbols a1, a2… label the damping strengthen modes.